\documentclass[%
reprint,
superscriptaddress,
 amsmath,amssymb,
 aps,
prl,
]{revtex4-1}

\usepackage{graphicx}
\usepackage{dcolumn}
\usepackage{bm}
\usepackage{color}

\begin{document}

\title{Quantum Spin Hall-like Phononic States in van der Waals Bilayers with Antiferromagnetic Ordering}

\author{W. H. Han}
\email{hanwooh@kaist.ac.kr, hanwooh@snu.ac.kr}
\affiliation{Department of Physics, Korea Advanced Institute of Science and Technology, Daejeon 34141, Korea}%
\affiliation{School of Law, Seoul National University, Seoul 08826, Korea}%

\author{K. J. Chang}
\affiliation{Department of Physics, Korea Advanced Institute of Science and Technology, Daejeon 34141, Korea}%

\date{\today}

\begin{abstract}
Here, we propose quantum spin Hall-like phononic states in van der Waals bilayer systems with antiferromagnetic ordering.
Antiferromagnetic ordering in bilayer systems with small interlayer interaction makes the total Chern number zero, where the Chern numbers for each layer are opposite to each other.
In bilayer CrI$_{3}$ where antiferromagnetic ordering has been experimentally demonstrated, we show that the quantum spin Hall-like states appear with spatially separated chiral edge modes.
\end{abstract}

\maketitle


Under broken time-reversal symmetry (TRS) conditions, many exotic states have emerged such as integer quantum Hall (IQH) states, fractional quantum Hall (FQH) states, and quantum anomalous Hall (QAH) states \cite{Klitzing1980,Thouless1982,Tsui1982,Haldane1988,Chang2013}.
In 2005, Kane and Mele showed that non-trivial topological states can be realized by spin-orbit coupling (SOC) effects even in TRS invariant systems, which is the origin of the concept of quantum spin Hall (QSH) states \cite{Kane2005Z,Kane2005Q}.
After successive works, the concept of topology has been extended to not only semimetals such as Dirac and Weyl semimetals \cite{Liu2014,Soluyanov2015,Deng2016,Armitage2018}, but also other fundamental particles, for example the photon \cite{Wang2009,Rechtsman2013,Lu2013}, magnon \cite{Zhang2013,Mochizuki2014,Chisnell2015}, and phonon \cite{Prodan2009,Kane2014,Susstrunk2016}.
For phonons, a quasi-particle from vibrations of atoms in solids, it has been revealed that when the TRS in phononic systems is broken by gyroscopic forces \cite{Wang2015,Nash2015} and spin-phonon interactions (SPI) \cite{Sheng2006,Zhang2010}, QAH-like states are induced with one-way edge modes.
However, few studies for QSH-like states have reported because the phonon does not have spin, in contrast to electrons.
Recently, several pesudo-spins have been proposed in hexagonal lattices and the Kekul\'{e} lattice \cite{Zhang2015,Liu2017_2,Liu2018}, showing the potential presence of QSH-like states in phononic systems.
It is therefore desirable to research QSH-like states in phononic systems for fundamental interest and practical applications based on phononics.

To go back to the simple model for QSH states proposed by Kane and Mele \cite{Kane2005Z}, it can be simply understood as two copies of the Haldane model with opposite spin Chern numbers to yield the total Chern number zero.
Although the phonon does not have spin, it is possible to imagine two identical phonons existing in real space with opposite Chern numbers.
In van der Waals (vdW) bilayer systems, upper and lower layers can be approximated to two identical copies in the way that interactions between layers are quite small without chemical interactions or surface reconstruction \cite{Geim2013,Duong2017}.
The remaining key point is how opposite TRS breaking terms are effectively implemented in the vdW bilayer systems.
Very recently, ferromagnetic ordering in two-dimensional (2D) materials such as CrI$_{3}$ and Cr$_2$Ge$_2$Te$_6$ have been reported in experiments \cite{Huang2017,Gong2017}, in which a magnetic anisotropy overcomes the Mermin-Wagner theorem \cite{Mermin1966}.
Moreover, it is interesting to note that in the bilayer CrI$_{3}$, magnetic ordering can be manipulated between ferromagnetic and antiferromagnetic ordering by electrostatic gate control \cite{Huang2018,Jiang2018,Klein2018}.
Since the antiferromagnetic ordering in the bialyer CrI$_{3}$ has opposite directions of magnetization, which determines the sign of SPI \cite{Sheng2006}, we can expect to realize QSH-like phononic systems consisting of two opposite QAH-like states.

In this Letter, we propose QSH-like phononic systems in vdW bilayer systems with antiferromagnetic ordering.
The QSH-like phononic state can be viewed as a combination of two copies of QAH states with the opposite direction of TRS breaking terms in each layer.
In this system, the two layers are considered to be pesudo-spins, and the opposite directions of the TRS breaking terms due to antiferromagnetic ordering act as pesudo-SOC interactions.
Through our newly developed program, we show that the QSH-like phononic states can exists in bilayer CrI$_3$ together with spatially separated chiral edge modes.
Based on our results, we discuss the difference between QSH states in an electronic system and the QSH-like phononic systems, and finally propose its potential applications which motivate future researches.

In order to calculate topological invariants in phononic systems under broken TRS conditions, a few Schr\"{o}dinger-like phononic hamiltonians have been proposed \cite{Susstrunk2016,Liu2017}.
Here, we employ a form of Schr\"{o}dinger-like phononic hamiltonian proposed by Liu \emph{et al.} \cite{Liu2017} to calculate topological invariants under TRS broken conditions, which can be expressed as
\begin{equation}
H_{k}\psi_{k} = \begin{bmatrix} 0 & iD_{k}^{1/2} \\ -iD_{k}^{1/2} & -2i\eta_{k} \end{bmatrix}  =  \omega_{k}\psi_{k} =  \omega_{k} \begin{bmatrix} D_{k}^{1/2} u_{k} \\ \dot{u_{k}} \end{bmatrix},
\end{equation}
where $D_{k}$, $\eta_{k}$, $\omega_{k}$, and $u_{k}$ represent dynamical matrices, TRS breaking terms, eigenvalues, and eigenvectors of the dynamical matrices, respectively.
A detailed explanation is given elsewhere \cite{Liu2017}.
This hamiltonian form is implemented in the phononTB program we have recently developed \cite{phononTB}, and topological invariants are calculated from periodic parts of the phononic wavefunctions ($\psi_{k}$).
In our phononTB program, force constants of a model system are constructed by Slater-Koster tight-binding parameters \cite{Slater1954} because  the $x$, $y$, and $z$ components for each atomic site can be viewed as three localized $p$ orbitals in three-dimensional space.
For the phonon dispersion curves of real materials, force constants calculated using the Phonopy program \cite{Togo2015} with first-principle calculations are directly read by the phononTB program to solve the Schr\"{o}dinger-like phononic hamiltonians.
In this work, to obtain the force constants of monolayer and bilayer CrI$_3$, 3 $\times$ 3 $\times$ 1 supercells were employed with the functional form proposed by Perdew, Burke, and Ernzerhof for the exchange-correlation (XC) potential \cite{Perdew1996} and Grimme's corrections \cite{Grimme2006}, as implemented in VASP \cite{Kresse1996}.

According to the previous study \cite{Liu2017}, an effective $k \cdot p$ equation for the monolayer honeycomb lattice is expressed as
\begin{equation}
H_{0}^{mono}(k) = k_{y}\tau_{z}\sigma_{x} - k_{x}\tau_{0}\sigma_{y},
\end{equation}
where $\sigma_{z}$ and $\tau_{z}$ represent the sublattice and valley index, respectively,
Here the Dirac velocity is set to be 1.
Among perturbations which anticommute with the hamiltonian, $\tau_{0}\sigma_{z}$ and $\tau_{z}\sigma_{z}$  correspond to inversion symmetry (IS) breaking and TRS breaking terms, which gives rise to trivial and non-trivial topological states, respectively \cite{Liu2017}.

First, we imagine a simple bilayer system consisting of two copies of a honeycomb lattice without any interlayer interactions (Fig. 1). 
To write the hamiltonian for the bilayer system, we introduce new Pauli matrices referring to the layer index, $\varepsilon_{z}$, in which 1 and -1 indicate an upper and lower layer, respectively.
Since there is no coupling between the two layers, they can be considered as independent systems, and therefore the total hamiltonian for the bilayer system is given as:
\begin{equation}
H_{0}^{bi}(k) = k_{y}\varepsilon_{0}\tau_{z}\sigma_{x} - k_{x}\varepsilon_{0}\tau_{0}\sigma_{0}.
\end{equation}
By using layer, valley, and sublattice indices, the SPI in the ferromagnetic and antiferromagnetic ordering in the vdW bilayer systems can be expressed as
\begin{equation}
\begin{split}
H_{T}^{FM} = \Omega_{z}\varepsilon_{0}\tau_{z}\sigma_{z} \\
H_{T}^{AFM} = \Omega_{z}\varepsilon_{z}\tau_{z}\sigma_{z}, 
\end{split}
\end{equation}
where $\Omega_{z}$ is the strength of SPI.
Since antiferromagnetic ordering has opposite magnetization directions in the upper and lower layers, the signs of the SPI should be opposite to each other while they are same in ferromagnetic ordering.

As shown in Fig. 1, at each K and K$'$ point, the two Dirac points coming from the two layers are degenerated.
When the ferromagnetic ordering is included, the Berry phases at K and K$'$ point are calculated to be $2\pi$ from the two independent copies of the QAH-like states.
Thus the total Chern number becomes 2. 
On the other hand, for the antiferromagnetic ordering in the bilayer, the Berry phases for the upper and lower layer cancel out at the K and K$'$ points, and then the total Chern number will be zero.
However, care should be taken that the Chern numbers projected onto each layer are not zero but opposite to each other.
This indicates that the bilayer with antiferromagnetic order has opposite layer Chern numbers.
This concept is very similar to an early idea of Kane and Mele except that TRS is explicitly broken in the antiferromagnetic ordering bilayer system.
We can expect that two one-way edge phonon modes in the opposite direction occur, but it is interesting to note that they should be spatially separated into each layer.
Therefore, we call it a QSH-like phononic state with spatially separated edge states.

We have shown that a bilayer with opposite TRS breaking terms will generate a QSH-like state if there is no interlayer interaction.
However, in many vdW layered systems, finite interlayer interactions between two layers exist even though their strengths are not significant.
In order to include interlayer interactions between 2D layers, we consider two types of interlayer interactions between an upper and lower layer: $A^{upper}$-$A^{lower}$ and $A^{upper}$-$B^{lower}$.
The former (latter) indicates interactions between the same (different) atoms in different layers.
The interaction strengths are denoted by $\gamma_{1}$ and $\gamma_{2}$, respectively, and within the same basis of the Eq.3 they are transformed into
\begin{equation}
\begin{split}
H_{inter}^{A-A} = \gamma_{1}\varepsilon_{x}\tau_{0}\sigma_{0} -  \gamma_{1}\varepsilon_{x}\tau_{y}\sigma_{z}  \\ 
H_{inter}^{A-B} = \gamma_{2}\varepsilon_{x}\tau_{z}\sigma_{y} - \gamma_{2}\varepsilon_{x}\tau_{x}\sigma_{x} .
\end{split}
\end{equation}
For the sake of simplicity, here we only consider interactions between same directions ($x$-$x$ or $y$-$y$), and then the second term in each $H_{inter}$ is neglected.
Since both terms include $\varepsilon_{x}$ causing phonon mixing between layers, all eigenmodes are separated into two eigenmodes,  with eigenvalues of -1 and 1 corresponding to bonding and anti-bonding states.
In Fig. 2(a), two degenerated linear dispersions near the K point are separated into two linear dispersions with a spacing of $\gamma_{1}$ or $\gamma_{2}$.
Therefore, the presence of interlayer interactions creates two modes from each intermixed layer.

Considering both the interlayer and TRS breaking interactions simultaneously, two interactions are competing for the gap opening.
As shown in Fig. 2(a), we consider ferromagnetic ($H_{T}^{FM}$) and antiferromagnetic ($H_{T}^{AFM}$) ordering type TRS breaking terms in $H_{0}^{bi} + H_{inter}^{A-A}$ and $H_{0}^{bi} + H_{inter}^{A-B}$.
Since all terms in the hamiltonian do not have interactions between the K and K$'$ points, the total hamiltonian of an 8 $\times$ 8 matrix can be split into two 4 $\times$ 4 matrices for each valley.
For the $H_{inter}^{A-A}$, we find that $H_{T}^{FM}$ induces a topological phase transition when $\Omega_{z}$ exceeds $\gamma_{1}$, while $H_{T}^{AFM}$ immediately results in non-trivial states with a finite gap size of $\Omega_{z}$.
However, the gap opening behavior for $H_{inter}^{A-B}$ is the opposite of that for $H_{inter}^{A-A}$.
Therefore, when considering both interlayer interactions, $H_{inter}^{A-A}$  and $H_{inter}^{A-B}$, the TRS breaking terms give rise to topological phase transitions only if their strengths are larger than the interlayer interactions.
The total phase diagram is summarized in Fig. 2(e).

From the effective $k \cdot p$ approach, it is found that a topological phase transition occurs when the TRS breaking interactions are larger than the interlayer interactions.
Here, using the phononTB program, we demonstrate a QSH-like phononic state in a simple phononic model consisting of two honeycomb lattices with Bernal (AB) stacking.
Instead of the interlayer interactions described above, distance-dependent interlayer interactions ($\gamma$) are employed, and a more detailed description is illustrated in Fig. 3(a).
When interlayer interactions are included, the Dirac cone at the K and K$'$ points becomes broken, and each valley has a berry phase of $2\pi$ and $-2\pi$, respectively. 
With opposite direction TRS breaking terms in each layer, which are larger than interlayer interactions, the total Chern number is calculated to be zero [Fig. 3(b)].
As shown in Fig. 3(c) and (d), in a zigzag nanoribbon configuration, two chiral edge states at the same edge propagate in the opposite direction, and they are separated into each layer.
In this case, it is expected that the total phonon transmission will be measured to be zero, but there is a non-zero local phonon transmission in each layer.
Through the model calculations, we confirm the presence of QSH-like phononic states with topologically protected edge states under antiferromagnetic ordering.

Finally, we apply the concept of QSH-like phononic states to bilayer CrI$_3$ which has been experimentally demonstrated  \cite{Huang2018,Jiang2018,Klein2018}.
Here we used experimental lattice constants measured from bulk phases at low temperature to obtain phonon dispersion curves \cite{Mcguire2015}.
As shown in Fig. 4(a), the higher phonon frequency of monolayer CrI$_3$ can be attributed to Cr atoms rather than I atoms due to their different atomic masses.
Among higher frequencies, we determined that the 20th and 21st modes around 6.4 THz consist of in-plane Cr displacements [Fig. 4(b)].
More interestingly, they have two Dirac points at the $\Gamma$ and K point, which have the potential to yield non-zero Chern numbers [Fig. 4(a)].
Although the accurate phonon frequencies of these modes depend on the XC functional forms employed, the two band touching points were already identified in the previous study \cite{Webster2018}.
Since the local magnetic moments are localized only on Cr atoms with the $z$ direction, we impose TRS breaking terms only for Cr atoms with $\Omega_{z}$ components.
The TRS breaking terms open Dirac cones between the 20th and 21st modes, giving rise to Chern numbers of 2 and -2, respectively [Fig. 4(a)].
Consequently, the appearance of two chiral edge states is observed by edge calculations, as shown in Fig. 4(c).

When two CrI$_3$ are stacked in AB stacking, the frequencies of the 20th and 21st modes become slightly split due to interlayer interactions.
We examined the ferromagnetic ordering and antiferromagnetic ordering type TRS breaking terms by manipulating the signs of the $\Omega_{z}$ components for the Cr atoms in each layer.
In the ferromagnetic ordering, the total Chern number was calculated to be 4, which is twice the Chern numbers for monolayer, and then two edge channels are formed in the upper and lower layers [Fig. 4(d)]. 
On the other hand, the anfierromagnetic ordering causes each layer Chern number to cancel out, and yields the total Chern number of zero.
Therefore, in bilayer CrI$_3$, it is shown that QSH-like phononic states are induced by the antiferromagnetic ordering, which results in opposite chiral edge modes [Fig. 4(e)].

Last but not least, we discuss differences between QSH-like phononic states and QSH electronic states.
First of all, since TRS is explicitly broken in QSH-like phononic states, the layer Chern number can be any integer, while QSH states have $Z_{2}$ invariants.
As shown in the above calculations, bilayer CrI$_3$ has four chiral edge states, and two of them have a different direction than the others.
Next, it is interesting to note that edge states in the QSH-like phononic states are not robust to backscattering because two states localized at different layers are not protected by symmetry.
In QSH states, it is well known that two spins propagating oppositely at the same edge do not interact each other due to presence of TRS \cite{Zhang2009}.
However, in QSH-like phononic states, two edge states at the same side but localized at different layers are hybridized when they cross, resulting in a finite band gap opening [Fig. 4(d) and (e)].
The size of the band gap depends on interlayer interactions, and thereby it vanishes when interlayer interactions are artificially decreased [Fig. 4(d) and (e)].
This behavior is similar to the hybridization between the surface states on different surfaces in very thin Bi$_2$Se$_3$ films, because two surface states in opposite sides are not protected by TRS \cite{Zhang2010_2,Neupane2014}.
This hybridziation gap may be useful in terms of device applications because within this hybridization gap there is no phonon transmission which is distinct from non-trivial edge states.

In summary, we have demonstrated QSH-like phononic states in vdW bilayer systems where two layers have different signs of magnetization.
Based on the model and \emph{ab initio} calculations, we suggest that bilayer CrI$_3$ with antiferromagnetic ordering can potentially realize the proposed QSH-like states.
It is expected that twice or zero conductance will be measured in magnetic vdW bilayer systems when the magnetic ordering is switched between ferromagnetic and antiferromagnetic ordering, and these findings may trigger future device applications such as spintronics and phononics.

\section{Acknowledgments}
\begin{acknowledgments}
This work was supported by Samsung Science and Technology Foundation under Grant No. SSTFBA1401-08.
\end{acknowledgments} 

\bibliography{QSHph_HAN}

\clearpage

\begin{figure}[ht]
\includegraphics[width=\columnwidth]{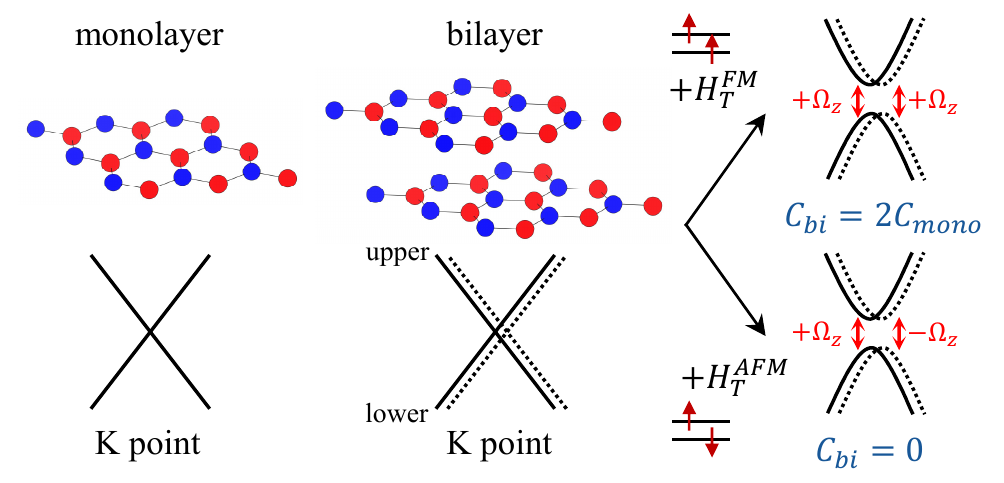}
\caption{\label{fig:1} A schematic diagram of the effects of TRS breaking terms in a vdW bilayer system without interlayer interactions.
The ferromagnetic ($H_{T}^{FM}$) and antiferromagnetic ($H_{T}^{AFM}$) ordering type TRS breaking terms cause two and zero Chern numbers, respectively.
} 
\end{figure}

\begin{figure}[ht]
\includegraphics[width=\columnwidth]{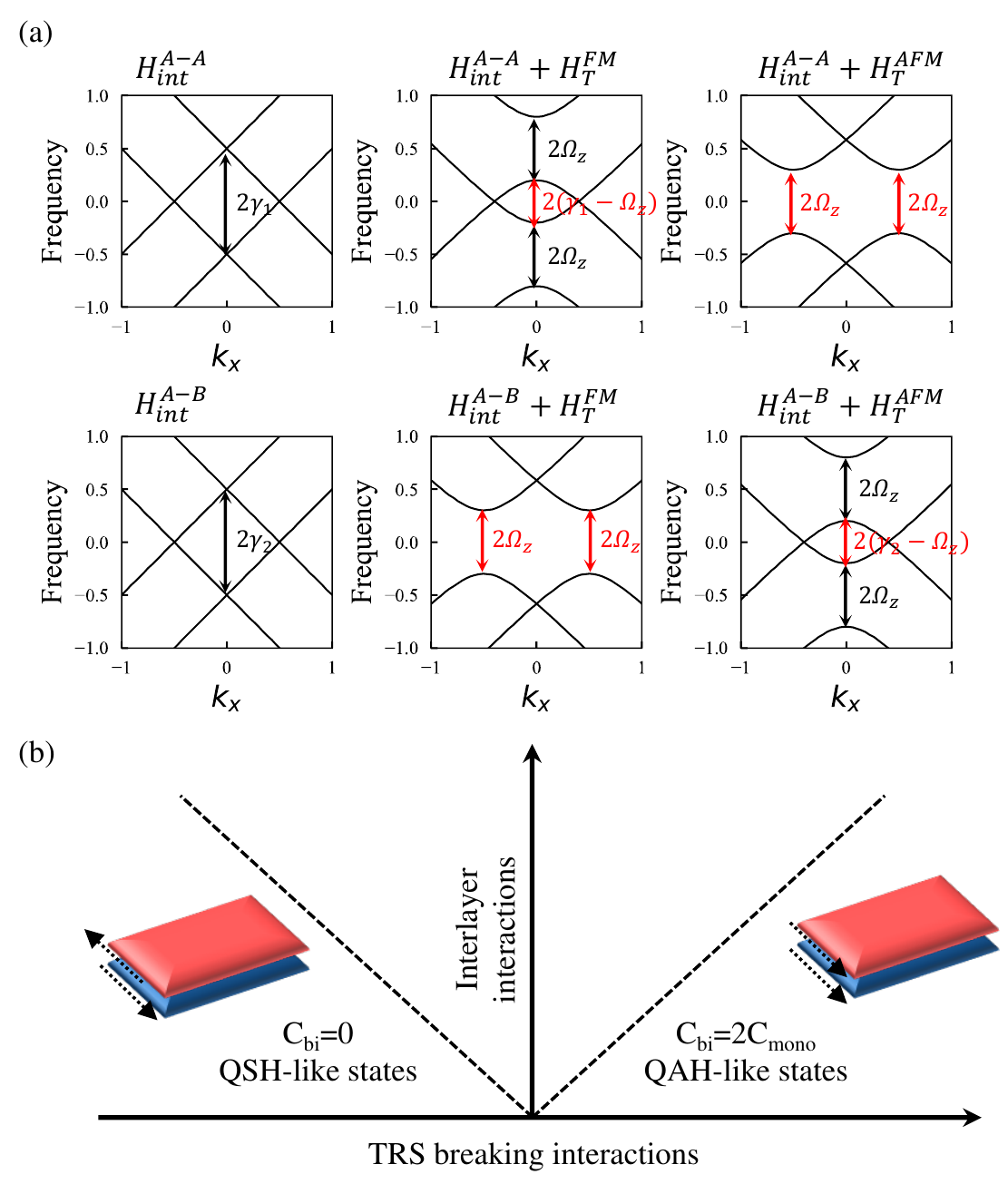}
\caption{\label{fig:2}
	(a) An effective $k \cdot p$ approach to consider interlayer interactions and TRS breaking terms.
	(b) A schematic phase diagram of QSH- and QAH-like states as functions of interlayer and TRS breaking interaction strengths.
} 
\end{figure}

\begin{figure}[ht]
\includegraphics[width=\columnwidth]{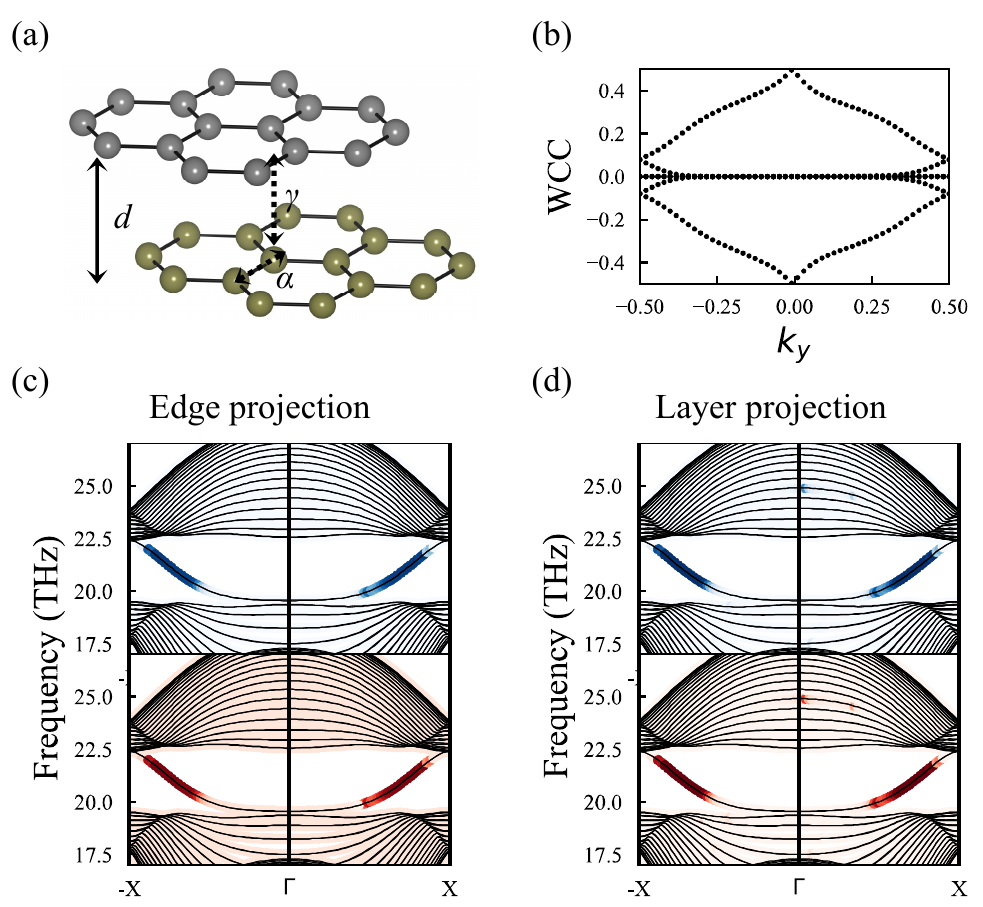}
\caption{\label{fig:3}
(a) The atomic structure for a bilayer honeycomb lattice model is drawn with intralayer ($\alpha$) and interlayer ($\gamma$) force constants. Here, we assume that the interlayer interaction is quite small as compared to the intralayer interaction.
(b) The evolution of the Wannier charge center (WCC) of the model system, including antiferromagnetic ordering type TRS breaking interactions.
(c) and (d) The edge states of the model system which are projected onto edges and layers, respectively, and red and blue colors indicate different edges and layers.
} 
\end{figure}

\begin{figure*}[ht]
\includegraphics[width=\textwidth]{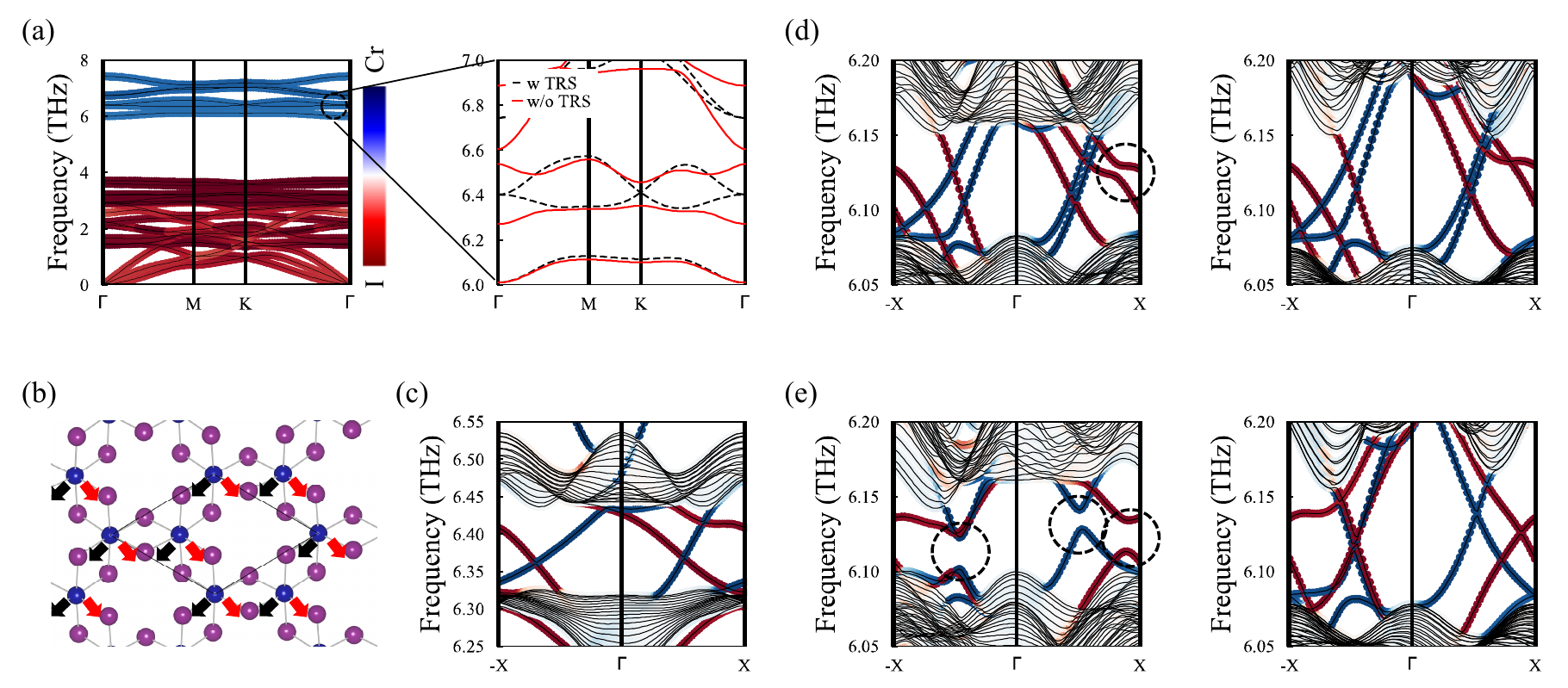}
\caption{\label{fig:4} 
(a) Atom-projected phonon dispersion curves for monolayer CrI$_3$. The blue and red circles indicate the atomic weights of the Cr and I atoms, respectively.
The magnified dispersion curve shows phonon frequencies of the 20th and 21st modes with and without TRS.
The two band touching points between the 20th and 21st modes become broken when breaking TRS.
(b) Atomic displacements of Cr atoms for the 20th and 21st phonon modes are drawn. The black and red arrows represent the 20th and 21st phonon modes, respectively. 
(c) The edge calculation results for monolayer CrI$_3$ when ferromagnetic ordering type TRS breaking terms are included. The blue and red circles represent edge-projections for right and left edges, respectively.
(d) and (e) The edge calculation results for bilayer CrI$_3$ with ferromagnetic and antiferromagnetic ordering type TRS breaking terms. The dotted circles indicate the finite gaps due to hybridization between two edge states in the same edge. The right figures are phonon dispersion curves when interlayer interactions are neglected artificially.
} 
\end{figure*}

\end{document}